\documentclass{aa501}
\usepackage{graphicx}
\usepackage{natbib}
\newcommand{\gone}{\gamma_1}
\newcommand{\gtwo}{\gamma_2}
\newcommand{\beq}{\begin{equation}}
\newcommand{\eeq}{\end{equation}}
\newcommand{\ben}{\begin{enumerate}}
\newcommand{\een}{\end{enumerate}}
\newcommand{\bei}{\begin{itemize}}
\newcommand{\eei}{\end{itemize}}
\newcommand{\bfig}{\begin{figure}}
\newcommand{\efig}{\end{figure}} 
\newcommand{\bea}{\begin{eqnarray}}
\newcommand{\eea}{\end{eqnarray}}

\newcommand{\bt}{\beta}
\newcommand{\gm}{\gamma}

\newcommand{\sigmaT}{\sigma_{\rm T}}
\newcommand{\tdiff}{t_{\rm diff}}
\newcommand{\qshock}{Q_{\rm s}}
\newcommand{\qhead}{Q_{\rm h}}
\newcommand{\qlobe}{Q_{\rm lobe}}
\newcommand{\gmin}{\gamma_{\rm min}}
\newcommand{\gmax}{\gamma_{\rm max}}
\newcommand{\gjet}{\Gamma_{\rm jet}}
\newcommand{\melec}{m_{\rm e}}
\newcommand{\diff}{{\rm d}}
\newcommand{\tcool}{t_{\rm cool}}
\newcommand{\td}{\tau}
\newcommand{\ggam}{\hat{\gamma}}
\newcommand{\rcore}{r_{\rm c}}
\newcommand{\dcore}{\rho_{\rm c}}
\newcommand{\phead}{p_{\rm h}}
\newcommand{\plobe}{p_{\rm lobe}}
\newcommand{\uhead}{u_{\rm h}}
\newcommand{\ulobe}{u_{\rm lobe}}
\newcommand{\ucmb}{u_{\rm c}}
\newcommand{\tstart}{t_{\rm start}}
\newcommand{\tjump}{\tau_{\rm D}}

\begin{document}

\title{Modelling the spectral evolution of classical double radio sources}
\author {K. Manolakou\thanks{On leave from the 
Department of Physics, University of Thessalonica} 
\and J.G. Kirk}
\offprints{J.G. Kirk}
\institute{Max-Planck-Institut f\"ur Kernphysik, Saupfercheckweg 1, 
D-69117 Heidelberg, Germany \\
\email{Dina.Manolakou@mpi-hd.mpg.de; John.Kirk@mpi-hd.mpg.de}
}
\date{Received \dots}
\abstract{
The spectral evolution of powerful double radio galaxies (FR II's) is 
thought to be determined 
by the acceleration of electrons at the termination shock of the jet, 
their transport through
the bright head region into the lobes 
and the production of the radio emission 
by synchrotron radiation
in the lobes.  
Models presented to date incorporate some of these processes  
in prescribing the
electron distribution which enters the lobes. We have extended these models to 
include a 
description of electron acceleration at the relativistic termination shock and a 
selection of 
transport models for the head region. These are coupled to the evolution of the 
electron spectrum in the
lobes under the influence of losses due to adiabatic expansion, by inverse 
Compton scattering on the
cosmic background radiation and by synchrotron radiation.
The evolutionary tracks predicted by this model are compared to observation 
using the power/source-size
({\em P-D\/}) diagram. We find that the simplest scenario, in 
which accelerated 
particles suffer 
adiabatic losses in the head region which become more severe as the 
source expands 
produces {\em P-D\/}-tracks 
which conflict with
observation, because the power is predicted to decline too steeply with 
increasing size.  Agreement with
observation can be found by assuming that adiabatic losses are compensated 
during transport between the
termination shock and the lobe by a re-acceleration process
distributed throughout the head region.
}
\maketitle

\section{Introduction} 

Powerful double radio galaxies or \lq classical doubles\rq\ (CDs) owe their name to the 
extended
(hundreds of kpc) lobes of radio emission they exhibit on opposite sides of the
parent galaxy.
\citet{fanaroffriley74} classified these sources as type II objects; they have 
luminosities $P_{178\,{\rm MHz}}>5 \times 10^{25}\,{\rm W\,Hz}^{-1}$
and are edge-brightened, with bright outer 
hotspots. 
It is universally agreed that the radio continuum of the CDs
is synchrotron radiation from relativistic electrons
and perhaps positrons. The standard scenario is that of a jet propagating from
the galaxy to the outer parts of the lobes, passing through a shock front at
the hotspots and subsequently filling a \lq cocoon\rq\ around the jet with radiating
particles --- see, for example, \citet{begelmancioffi89,peacock99}.

\citet{falle91} considered radio sources as expanding 
bubbles that are fed by supersonic jets and drive a bow shock into a
radially stratified external medium. He showed that the jet length and the 
bow shock grow in a self-similar way for external atmospheres in which the 
density
drops off more slowly than $1/r^2$ from the centre of the galaxy. 
Subsequently, \citet{kaiseralexander97} showed that the 
cocoons also expand self-similarly. Based on this result, \citet{kaiseretal97}
developed an analytical model for the spectral evolution of FR~II sources as
a function of redshift, jet power and 
the scaling of the external density profile. 
They assumed that electrons with an initial power-law distribution in energy are
continuously injected into the plasma immediately downstream of the shock
front that terminates the jet.
A key point of their treatment is that the pressure in this region equals
that which drives the bow shock into the external medium on the axis of the
source. This results from the fact that the jet is confined by the pressure of
the cocoon \citep{begelmancioffi89}, so that the working surface over which its thrust is distributed
expands self-similarly, along with the bow shock and cocoon. 
Particles undergo adiabatic losses on moving from the terminal shock front
into the 
main part of the cocoon, whose pressure is lower than that at the working 
surface by a constant
factor (taken as $\approx16$ by \cite{kaiseretal97}).
They then undergo synchrotron and inverse Compton losses, as well as adiabatic
losses as the lobe of the source expands. Using this model, 
\cite{kaiseretal97} computed evolutionary tracks in the 
power-linear size ({\em P-D}) plane. These display 
decreasing power as the size increases, in agreement with observations, 
which indicate a deficit of large luminous sources.

However, \cite{blundelletal99} noted that there is no
evidence that the size of the working surface is proportional to the 
size of the source. Instead, the hot-spots of all known FR~II sources appear to 
be a few kiloparsecs in diameter. This has a profound influence on the 
predicted evolutionary tracks, because it implies that 
the adiabatic losses suffered 
upon moving into the cocoon by 
particles accelerated at the termination shock are not constant, but increase 
strongly with the age of the source.
\citet{blundelletal99} modelled the spectral
evolution of CDs including this effect by 
prescribing the spectrum of electrons 
entering the cocoon to be a broken power-law distribution. They, too, found 
evolutionary tracks that
agree with the observed lack of large, luminous radio sources, as well as
with several other properties of the samples they investigated.

In this paper we present computations of evolutionary tracks based on the 
picture adopted by \citet{blundelletal99}. However, instead of prescribing the 
electron distribution that enters the cocoon, we assume that electrons are 
accelerated by the first-order Fermi process at the termination shock and  
then 
propagate through the hot spot region into the cocoon 
or \lq lobe\rq\  
according to one of two 
models, which we designate case A and case B. 
In case~A, the full adiabatic energy loss corresponding to the 
age dependent pressure difference between the hot spot and cocoon
is applied. This results in $P$--$D$ tracks which are in conflict with the 
observations. One possible way out of this problem is to assume that 
a re-acceleration process occurs after the initial encounter with 
the termination shock. Such a process is indicated 
independently by investigations of the spectra of individual hot-spots
and of optical synchrotron emitting jets
\citep{meisenheimeretal96,meisenheimeretal97,perleyetal97,wagnerkrawczynski00}. 
This motivates our case~B model, in which we assume that the 
adiabatic losses between the termination shock and the lobe
are compensated by a re-acceleration process during propagation 
through the head region.
In both cases 
electrons are carried through the high-loss region by fluid elements 
whose residence time is distributed according to a 
specified transport equation. During 
this time they suffer synchrotron and inverse Compton losses. 
After entering the cocoon, we 
follow the particle distribution as it cools and radiates and compute 
evolutionary tracks as well as snapshots of the spectrum at different ages. 

The paper is organised as follows: in Sect.~2 we summarise the hydrodynamic 
picture which we adopt for the source evolution. In Sect.~3 our treatment of 
Fermi acceleration is described, the transport model is specified and 
the kinetic equation obeyed by particles in the cocoon is formulated. 
Section~4 
is devoted to a series of tests and special cases to illustrate the way in 
which the transport and acceleration models influence the 
particle distribution 
in the hot spot and the lobe. Our main results on spectral evolution are 
presented in Sect.~5, where we conclude that the observed data 
exclude case~A i.e., they 
can be 
explained only by a model such as case~B, 
that includes re-acceleration of electrons after their 
encounter with the termination shock. 
This conclusion and our results on source 
spectra at different ages are discussed and compared with previous work in 
Sect.~6. 

\section{Hydrodynamic Model}
\label{hydro} 
\begin{figure}
\resizebox{\hsize}{!}
{\includegraphics{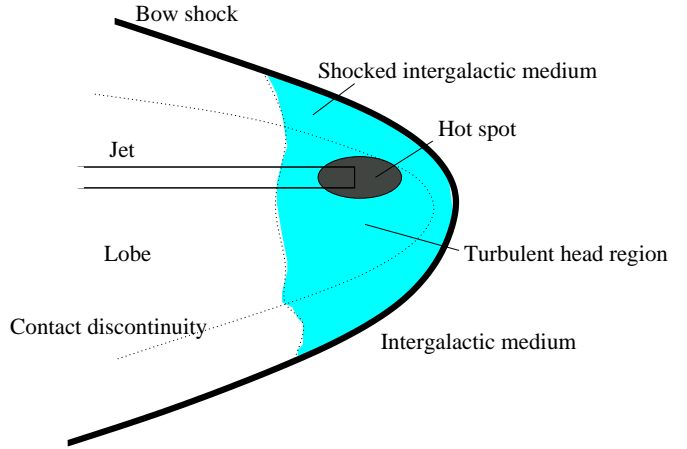}}
\caption{\protect\label{cartoon}
A cartoon of the interaction of the intergalactic medium and a jet of an 
FR~II radio galaxy, showing the principal components of 
the spectral evolution model. A jet of constant Lorentz factor and power is
decelerated in a {\em hot spot} (whose size is independent of source
age) converting part of its energy into relativistic electrons and part into
magnetic field. The electrons propagate according to a prescribed transport
model [parameterised by $\alpha$ and $\tau$---see 
Eq.~(\protect\ref{modulation})] from the hotspot 
into the turbulent head region. The hot spot moves around inside 
the head region; averaged over time, the jet thrust is balanced by    
the pressure in the head region, which expands self-similarly along with
the bow shock and lobe. Synchrotron losses are important in the (constant)
magnetic field of the hot spot, but
adiabatic losses suffered in moving into the head region and later into the
lobe may be compensated by a re-acceleration process acting in
either the hot spot or the head region. Once in the lobe, time-dependent
synchrotron, inverse Compton and adiabatic losses determine the 
spatially integrated electron spectrum. 
}
\end{figure}

The standard model for the CD radio sources comprises a central object (AGN) 
from which emanate two jets in opposite directions, embedded in a cocoon or 
lobe of diffuse emission. The jets themselves terminate in regions of intense 
emission, 
called the hotspots
or working surfaces, after which the shocked jet material flows into the
cocoon. 
This material drives a bow shock into the external medium, from which it is
separated by a contact discontinuity.
A cartoon of the model is shown in Fig.~\ref{cartoon}.

Following \citet{blundelletal99}, we
assume that the radial profile of the external (intergalactic) medium does not
vary with 
redshift and scales with distance $r$ from the centre of the host 
galaxy as
\beq
\rho={\dcore}(r/{\rcore})^{-\bt}
\label{denprof}
\eeq
\citep{garringtonconway91,mulchaeyzabludoff98}.
Typical values for the above parameters are: 
${\dcore}=1.67 \times 10^{-23}\,$kg 
m$^{-3}$, ${\rcore}=10\,$kpc 
and $\bt=1.5$.

During the lifetime of the source, we assume that the 
power $Q_0$ carried by the
jet and its Lorentz factor $\gjet$ remain constant.
The jet terminates at a shock front which we identify as the {\em primary
hotspot\/} embedded in a high-pressure region 
at the end of the lobe  
which, following the terminology of \citet{blundelletal99}, we call the {\em 
head region}.
The exact position of the termination shock within this region
is assumed to vary in the manner 
proposed by \citet{scheuer82} in the \lq dentist's
drill model\rq. 
Averaged over time, the thrust of the jet
is distributed over the head region, which itself expands with the 
rest of the source. However, the size of the primary hotspot is independent of
source age. Although its ram pressure substantially exceeds the average
pressure in the head, simulations indicate that it does not 
advance much more rapidly \citep{coxetal91}, because it quickly
encounters material of higher than average density. Consequently, the 
pressure  ${\phead}$ immediately downstream of the termination shock can be estimated 
as
the jet thrust, $Q_0/c$, divided by the area, $A$, 
of the primary hotspot 
itself. Assuming the jet is relativistic and 
the magnetic field is amplified to reach equipartition with
the thermal energy density, 
we may express the (constant) magnetic energy density in the 
hotspot, $u_{\rm h}$, in terms of 
the jet power as follows:
\begin{equation}
u_{\rm h} \sim {3 \over 2} {\phead} = {3 \over 2} {Q_0\over c A}
\enspace.
\label{uh}
\end{equation}

\citet{falle91} and \citet{komissarovfalle98}, found solutions in
which the size of a radio 
source increases in a 
self-similar way with time, with the total length scaling as:
\begin{equation}
D(t)=2 c_1 \left({t^3 Q_0 \over  {\dcore}\rcore^\beta}\right)^{1/(5-\beta)}
\enspace.
\label{ssize}
\end{equation}
From the measured expansion speeds of powerful sources one concludes that
$c_1\approx1.8$ \citep{blundelletal99}. 

In contrast to the constant volume and magnetic field of the hotspots,
the lobes of the source expand self-similarly.
Applying the jump conditions at the external bow shock and using
Eq.~(\ref{ssize}), 
gives for the pressure of the downstream 
plasma immediately behind 
the bow shock: 
\bea
 p_{\rm d}(t) &=& {18 \over \Gamma_{\rm x} +1} {c_1^{(2-\beta)}\over (5-\beta)^2}
 \left(\dcore\rcore^\beta\right)^{3/(5-\beta)}\times
\nonumber\\
&&Q_0^{(2-\beta)/(5-\beta)}
t^{-(4+\beta)/(5-\beta)}
\label{pd}
\eea
\citep{kaiseralexander97}
where $\Gamma_{\rm x}$ is the ratio of specific heats of the external gas, which we
take to be $5/3$.

The pressure in the lobes of the 
source, ${\plobe}$, is somewhat lower than this value, depending on the axial
ratio of the source. In this way the pressure gradient 
is provided to drive a 
high speed
back-flow from the head towards the core of the galaxy \citep{liuetal92}.
Assuming, in addition, 
that the magnetic field is tangled and that equipartition is 
also attained in the lobes,    
leads to the conclusion that the magnetic energy density in the lobe, ${\ulobe}$, 
scales as:
\begin{equation}
{\ulobe}(t)={\uhead}\left({ t \over t_0}\right)^{-a}
\label{ul}
\end{equation} 
where $a=(4+ \beta)/{(5-\beta)}$ and $t_0$ is the time 
at which the size of the head region was comparable to that of the primary hot
spot:
\bea
t_0&=&\left[{3 \over \Gamma_{\rm x}+1} {{c_1}^{2-\beta} \over
(5-\beta)^2} c A\right]^{1/a} 
\left({{\dcore}{\rcore}^{\beta} \over Q_0}\right)^{3/(4+\beta)}
\enspace.
\eea

\section{Particle acceleration and transport}
\label{accel}
We assume that the first-order Fermi process operates at the termination shock 
in the jet and accelerates there electrons (and possibly 
positrons) into a distribution
which can be described by a power-law in energy. 
As they move away from the shock these electrons radiate and 
form the primary hot spot within the head region. We can consider 
this process as an injection of electrons at a rate $\qshock(\gamma)$ into the
plasma behind the termination shock:
\beq
\qshock(\gamma)=
\left\lbrace \begin{array}{ll}
q_0 \gm^{-p}& \textrm{for $\gmin \le \gamma \le \gmax$ } \\
        &       \\
0 & \textrm{otherwise.}
\end{array}
\right.             
\label{sourcef}
\eeq
If the jet Lorentz factor $\gjet$ is of the order of 10, we can choose
$2<p<2.3$  \citep{kirketal00,achterbergetal01}. The upper and lower 
limits on the Lorentz factor 
$\gmax$ and $\gmin$ are not available from theory. It is 
plausible to assume $\gmin\approx\gjet$, but $\gmax$ should be determined by a
local balance of the acceleration and cooling rates. Fortunately, our models
are not sensitive to this parameter, since synchrotron cooling within the head
region will in any case determine the effective 
maximum energy of those particles which
enter the lobe.
The fraction of the jet power, 
$ \eta Q_0$, (with $\eta<1$) which is transfered 
into accelerated particles at the termination shock is
$\int \gamma\melec c^2 \qshock\diff\gamma$. By means of Eq.~(\ref{sourcef}) 
this 
gives for $q_0$:
\beq
q_0={\eta Q_0 \over m_e c^2} (p-2) (\gmin^{2-p}-\gmax^{2-
p})^{-1}
\enspace.
\label{normqo}
\eeq

After injection into the primary hot spot, electrons
are transported by the turbulent motions of the plasma through the head,
and eventually reach the lobes. 
During this time, which we assume is short
compared to the lifetime of the source, they are subject to
synchrotron losses in
the relatively strong magnetic field generated behind the shock
as well as inverse Compton losses by scattering off the cosmic background 
radiation. 
The effect which this has on the power-law distribution injected at the
termination shock depends on the distribution of the \lq escape time\rq\
i.e., on the probability an injected particle has of leaving the region of
strong field after a certain time. 

This distribution can be modelled in
several ways. The simplest approach would be to assume each particle escapes
after a fixed time, corresponding, for example, 
to the time taken for a fluid element to
cross the head region at a given speed. 
Upon reaching the lobes, the particles would then have 
the distribution computed by \citet{kardashev62}, which vanishes
above a cut-off energy at which the escape time equals the cooling time.
However, the turbulent appearance of the hot spot and head regions indicates
that not all elements of shocked plasma will take the same time to
escape into the lobes. 
At the next level of sophistication, this can be modelled in a
diffusion picture --- a plasma blob, and along with it accelerated
particles, performs a random walk through the head region. 
A convenient formalism, which has been widely used to describe transport 
in turbulent fluids and other random media \citep{bouchaudgeorges90} 
is given by the 
theory of continuous time random walks
(see \citet{ragotkirk97} and references therein). Diffusion in space 
results if a sequence of jumps of fixed length $\sigma$ 
in a random direction is
performed, where each jump takes the same time ${\tjump}$.
The advantage of this formalism is that by changing the waiting time 
distribution between
jumps it is easily generalised to 
include non-diffusive or \lq anomalous\rq\ transport characterised by a single 
transport parameter $\alpha$. The mean square distance 
$\left<\Delta r^2\right>$ of a
particle from its starting point is linearly proportional to the 
elapsed time only in the special case of diffusive transport $\alpha=1$; in
general one has $\left<\Delta r^2\right>\propto t^\alpha$ with $0<\alpha<2$. 
Sub-diffusive transport ($\alpha<1$) occurs when the motion of the fluid
elements is confined in some way  --- for example to forwards and backwards
motion in a channel, which itself diffuses sideways. Supra-diffusive transport,
on the other hand, corresponds to a more \lq free-streaming\rq-like 
motion, less
impeded by the random eddies. 
The spatial propagator for the generalised transport 
process in spherical geometry is
\bea
P(r,t)&=&
 \left[
{3\over2}\left({\alpha{\tjump}\over t}\right)^{\alpha}
\left({\sigma\over r}\right)^{2(1-\alpha)}\right]^{3/[2(2-\alpha)]}
{\sqrt{2-\alpha}\over\sigma^3\pi^{3/2}}
\nonumber\\
&&
\exp\left\lbrace -(2-\alpha)\left[{3r^2\over2\sigma^2}\left(
{\alpha{\tjump}\over t}\right)^\alpha\right]^{1/(2-\alpha)}\right\rbrace
\label{propgeneral}
\eea
where ${\tjump}$ is a characteristic timescale for the jumps.

During propagation, 
the rate at which an electron of energy $\gamma(t)\melec c^2$ loses 
energy is described by the equation:
\begin{equation}
{ \diff \gamma \over\diff t}=-g \gamma^2, \qquad g={4 \over 3}{\sigma_{\rm T} 
\over 
{\melec c}}({\uhead} + {\ucmb})
\label{gdoth}
\end{equation}        
where $\sigma_{\rm T}$, $\melec$ 
and $c$ are the Thomson 
cross section, the electron rest mass and the speed of light, respectively.
The energy density ${\ucmb}$ of the cosmic background radiation field is given
in terms of an effective magnetic field $B_{\rm c}$ by
\begin{equation}
{\ucmb} ={B_{\rm c}^2  \over {2 \mu_0}}, \qquad B_{\rm c}=0.318 
({1+z})^2 \,{\rm nT}
\label{uc}
\end{equation}
and is taken to be constant during the lifetime of a radio source.   
Equation~(\ref{gdoth}) may be integrated to give
$\gamma^{-1}(t)=\gamma^{-1}(t')+(t-t')g$. These loss processes may be included
on the right-hand side of Eq.~(\ref{propgeneral}), 
leading to a composite propagator
$\bar{P}({\bf x},t,\gamma,\gamma')$ that expresses the probability of finding a
particle with Lorentz factor $\gamma$ at ${\bf x}$ and $t$, given that it
started at the origin (${\bf x}=0$, $t=0$) with Lorentz factor
$\gamma'$:
\bea
\bar{P}({\bf x},t,\gamma,\gamma')&=&P({\bf x},t)
\delta\left[\gamma-{\gamma'\over1+tg\gamma'}\right]
\enspace.
\label{qhead2}
\eea

The distribution $N_R(\gamma)$ in Lorentz factor of the total number of particles contained in
a sphere of radius $R$, at the centre of which particles are injected
according to Eq.~(\ref{sourcef}) starting at time $t_{\rm start}$ is
\bea
N_R(\gamma)&=&
4\pi\int_0^R\!\diff r\, r^2\!\int_\gamma^\infty\!\!\diff\gamma'
\int_{t_{\rm start}}^t\!\!\!\!\diff t'
\nonumber\\
&&\bar{P}(r,t-t',\gamma,\gamma')\qshock(\gamma')
\enspace.
\eea
The two timescales that determine $N_R$ are the 
transport timescale (a generalisation of the diffusion time)
\bea
\tdiff&=&
(2-\alpha)^{(1-\alpha)/\alpha}
\left({3(2-\alpha)R^2\over2\sigma^2}\right)^{1/\alpha}\alpha{\tjump}
\eea
and the cooling time, defined for a particle
of Lorentz factor $\gmin$ as $\tcool=1/(g\gmin)$. If these are both 
short compared to the age of the source, it is reasonable to assume the 
transport process into the lobes reaches a steady state, found by setting
$t_{\rm start}\rightarrow-\infty$:
\bea
N_R(\gamma)&=&\left\lbrace
\begin{array}{l}
0\qquad\textrm{for $\gamma>\gmax$}\\
 \\
{4\pi q_0\over g\gamma^2}\int_\gamma^{\gmax}
\! \diff\gamma'\gamma'^{-p}\int_0^R\diff r\, r^2
P(r,s)
\\
\qquad\textrm{for $\gmin<\gamma<\gmax$}\\
 \\
{4\pi q_0\over g\gamma^2}\int_{\gmin}^{\gmax}
\! \diff\gamma'\gamma'^{-p}\int_0^R\diff r\, r^2
P(r,s)
\\
\qquad\textrm{for $\gamma<\gmin$}
\end{array}
\right.
\eea
where $s=(\gamma'-\gamma)/(g\gamma\gamma')$.
The kinetic equation satisfied by $N_R$ reads
\bea
{\partial N_R\over\partial t}-{\partial \over\partial \gamma}\left(g\gamma^2
  N_R\right)
&=&\qshock(\gamma)-\qhead(\gamma)
\label{kinhead}
\eea
where $\qhead(\gamma)$ is the rate at which particles leave the
sphere of radius $R$ (and so enter the lobe).
In the steady state, it is straightforward to solve Eq.~(\ref{kinhead})
for $\qhead$. The result is conveniently expressed as a modulation function
$M$ describing the effects of cooling on the incident distribution $\qshock$,
and defined as
\bea
\qhead(\gamma)&=&q_0\gamma^{-p}M(\ggam,\rho_{\rm d},p,\td)
\label{mdef} 
\eea
where we have also defined parameters 
representing the Lorentz factor: $\ggam=\gamma/\gmin$, the reciprocal of the 
dynamic range of the incident distribution: $\rho_{\rm d}=\gmin/\gmax$ and the 
dimensionless ratio of the transport time to the cooling time of a particle at
$\gmin$: $\td=\tdiff/\tcool$.
From Eq.~(\ref{kinhead}) we find
\bea
M(\ggam,\rho_{\rm d},p,\td)&=&
\left\lbrace
\begin{array}{l}
0\qquad\textrm{for $\ggam > 1/\rho_{\rm d}$}\\
 \\
{2\over
\sqrt{\pi}}\int_{x_1}^\infty\!\diff x\,
x^{1/2}\left[1-{\td\ggam\over x^{(2-\alpha)/\alpha}}
\right]^{p-2}
e^{-x}\\
\qquad\textrm{for $1<\ggam<1/\rho_{\rm d}$}\\
 \\
{2\over
\sqrt{\pi}}\int_{x_1}^{x_2}\!\diff x\,
x^{1/2}\left[1-{\td\ggam\over x^{(2-\alpha)/\alpha}}
\right]^{p-2}
e^{-x} \\
\qquad\textrm{for $\ggam<1$}
\end{array}
\right.
\label{modulation}
\eea
where 
\bea
x_1&=&\left[{\ggam\td\over(1-\rho_{\rm d}\ggam)}\right]^{\alpha/(2-\alpha)}
\nonumber\\
x_2&=&\left[{\ggam\td\over(1-\ggam)}\right]^{\alpha/(2-\alpha)}
\enspace.
\eea

Note that transport from the primary hot spot 
into the lobes results from the diffusion of
plasma elements in a turbulent environment. The mean free paths involved are
much larger than those estimated for individual energetic electrons, which are
usually taken to be of the order of the gyro radius. The turbulent mean free
path is also independent of the particle energy.  
However, this expression still does not account either for 
adiabatic losses or for 
any acceleration process that might occur between leaving the termination
shock and entering the lobe. 

The effect of adiabatic losses is to reduce the energy of each particle by 
a factor given by the ratio of the pressures before and after the expansion
[see, for example, \citet{scheuerwilliams68}]. 
In the case of a transition from the pressure in the primary hot spot to the
pressure in the lobe we have $\gamma\rightarrow({\ulobe} /{\uhead})^{1/4}\gamma$.
As a result, the slope of the power-law part of the distribution 
remains unchanged, but the upper and lower limits shift downwards in energy.
Of course, the situation is more complicated if, simultaneously,  
radiation losses are important. Our treatment of 
losses in the lobes, where an explicit time dependence of the energy density
is available, takes this explicitly into account. Qualitatively, however,
the form of the spectrum should not depend on the sequence 
in which these processes operate. In treating propagation through 
the head region we impose the adiabatic losses after the radiation losses, 
bearing in mind that the 
values derived for quantities such as the diffusion time $\tdiff$ 
may depend somewhat on this choice. 
Thus, assuming constant injection at the termination shock
and requiring particle number conservation, we find for 
the function describing the injection of electrons into the lobes:
\bea
\qlobe(\gamma,t)&=&k\qhead(k\gamma,t)
\label{qlobea}
\eea
where $k(t)=\left[{\uhead} /{\ulobe}(t)\right]^{1/4}$.

Re-acceleration by multiple encounters with shock fronts has been considered
in many papers, notably
\citet{spruit88,achterberg90,anastasiadisvlahos93,schneider93,popemelrose94,marcowithkirk99,gieselerjones00}.
In our model, we distinguish between the termination shock that forms
the primary hot spot and a sequence of subsequent, weaker shocks through which
the plasma passes before emerging into the relatively inactive lobe region. 
The theory of diffusive shock acceleration predicts that 
a single encounter with a weak shock will not
change the power-law index of an incoming distribution produced at a strong
shock --- the effect is merely to increase the amplitude of the distribution,
and to produce a steep power-law tail above the upper 
cut-off energy of the incoming
particles. Passage through many such shocks, however, tends also to harden a
preexisting power-law if the ratio of the escape time to the time between 
shock encounters is large. Shock drift acceleration, on the other hand, 
produces an
effect which is similar to an adiabatic compression
\citep{begelmankirk90,anastasiadisvlahos93}, in that the power-law index is
unchanged, but the distribution is shifted and slightly modified around the
cut-off energies. 
{\bf The effect of a highly turbulent environment comprising of a mixture
of different types of accelerators has been studied by \citet{manolakouetal99}
who conclude that the level of turbulent activity has at most a minor
impact on a pre-existing power-law index, and contributes only
to the efficiency of the process.}
Given the variety of possible effects, it is clearly 
a difficult observational challenge to distinguish between the different 
mechanisms. In this paper, we do not adopt a specific acceleration
model, but simply assume that when re-acceleration is important a distribution
emerges into the lobe which can be described as a power-law above 
a lower cut-off energy, modified by
synchrotron and inverse Compton losses at high energy. This corresponds
precisely to a modification of the effectiveness of the adiabatic losses
described by the parameter $k(t)$. If we now interpret the quantity $\eta$ 
as that fraction of the jet power which goes 
into energetic particles which enter the lobe, then the effects of
re-acceleration correspond simply to setting $k(t)\equiv1$ in 
Eq.~(\ref{qlobea}).

\subsection{The evolution of the electron distribution in the lobes}

The spatially integrated distribution $N(\gamma,t)$ of particles in the
lobes, is governed by a kinetic equation similar to Eq.~(\ref{kinhead}):
\beq
{\partial N \over {\partial t}} +{\partial \over {\partial \gamma}}(\dot
{\gamma} N) = \qlobe(\gamma, t)\enspace.
\label{diflobe}
\eeq
However, in this case the loss term is more complicated, involving adiabatic
losses as well as synchrotron and Compton losses, the first two 
of which have an explicit time-dependence:
\beq
\dot{\gamma} = -b {\gamma \over t} - b_{\rm ic} {\gamma}^2 - 
b_{\rm s}{{\gamma}^2        
\over t^{a}}\enspace.
\label{gdot}
\eeq
The three terms on the right hand-side of Eq.~(\ref{gdot}) represent
(i) adiabatic losses, with $b=a/4$ according to Eq.~(\ref{ul}), (ii) losses by 
inverse
Compton scattering, with 
$b_{\rm ic}=({4/ 3})({\sigmaT/\melec c}){\ucmb}$ a constant, and 
(iii) synchrotron losses in the adiabatically decreasing magnetic field
of the lobe [see eq.~(\ref{ul})]: $b_{\rm s}=({4/3})({\sigmaT/ \melec c}){\uhead} t_0^a$. 

Following the method of characteristics, Eq.~(\ref{diflobe}) 
may be rewritten in terms of the derivative of $N$ 
along a family of curves $\chi(\gamma,t)$ describing 
the evolution of individual particles and given by integrating
Eq.~(\ref{gdot}):
\beq
\chi(\gamma,t)= {1 \over t^b}
\left( {1 \over \gamma} - {b_{\rm ic} t \over 1-b}-
{b_{\rm s} t^{1-a} \over 1-b-a}\right)
\enspace.
\label{chi}
\eeq
Then $N(\chi,t)$ 
satisfies the ordinary differential equation
\beq
{\diff N \over \diff t}-
\left[ {b \over t}+ 2 \gamma (b_{\rm ic}+b_{\rm s} t^{-a})\right]N=
\qlobe(\gamma,t),
\label{odef}
\eeq
in which $\gamma$ is considered a function of $\chi$ and 
$t$ according to Eq.~(\ref{chi}).

Eq.~(\ref{odef}), with $\qlobe$ given by Eq.~(\ref{qlobea}),
may be integrated numerically, the upper and lower limits of 
integration being
the observation time, $t$ and the 
earliest time $t_{\rm i}(\gamma,t)$
at which the source injected particles into the lobes that subsequently
cooled to
the chosen Lorentz factor $\gamma$ at time $t$.
The initial condition is $N(\chi,t_{\rm i})=0$. 

The injection time $t_{\rm i}$ 
is a function of the Lorentz factor, $\gamma$ and the 
observing time, $t$. 
It is determined not only by the cooling rate, but also, at least
potentially, by the time $\tstart$ at which the source started its
activity. 
This limit
is important if particles injected into the lobe at time ${\tstart}$ with 
$\gamma_{\rm i}<\gmax/k({\tstart})$ can cool to $\gamma$ at time $t$. From 
Eq.~(\ref{chi}), this situation arises when
\bea
{1 \over \gamma_{\rm i}}&=& {\tstart} 
\left({b_{\rm ic} \over 1-b}+
{b_{\rm s} {\tstart}^{-a} \over 1-b-a}\right)+{\tstart}^b \chi(\gamma,t)
\nonumber\\
&<&{k({\tstart})\over \gmax}
\label{gilt}
\eea
in which case we set $t_{\rm i}={\tstart}$. 
If, on the other hand, $\gamma_{\rm i} >\gmax/k({\tstart})$, then 
$t_{\rm i}$, is found from the relation:
\beq
{k(t) \over \gmax}= t_{\rm i} 
\left({b_{\rm ic} \over 1-b}+
{b_{\rm s} t_{\rm i}^{-a} \over 1-b-a}\right)+t_{\rm i}^b \chi(\gamma,t)
\eeq
In summary, the injection time, $t_{\rm i}$ lies in the range ${\tstart} < t_{\rm i} < 
t$, if 
$\gamma_{\rm i} > \gmax/k({\tstart})$ and equals ${\tstart}$  if 
$\gamma_{\rm i} < \gmax/k({\tstart})$.

Finally, the specific power $P_\nu(t)$ (power per unit 
frequency) is found by integrating the product of $N(\gamma,t)$ and the 
emissivity of a single electron. Adopting the delta-function approximation we 
have:
\begin{equation}
P_{\nu}(t)={1 \over 4 \pi}\int_{1}^{\infty}a_0 \gamma^2 B_{\rm L}(t)^2
\delta(\nu-a_1B_{\rm L}(t)\gamma^2)
N(\gamma,t) d \gamma 
\label{power}
\end{equation}      
where $B_{\rm L}(t)$ is the magnetic field in the lobe at time $t$, 
$a_0=1.6 \times 
10^{-14}{\rm W\,T^{-2}}$ and $a_1=1.3 \times 10^{10}{\rm Hz\,T^{-1}}$. 

\section{Tests and special cases}

\subsection{Cooling in the head region}

\begin{figure}
\resizebox{\hsize}{!}
{\includegraphics[bb=18 151 574 697]{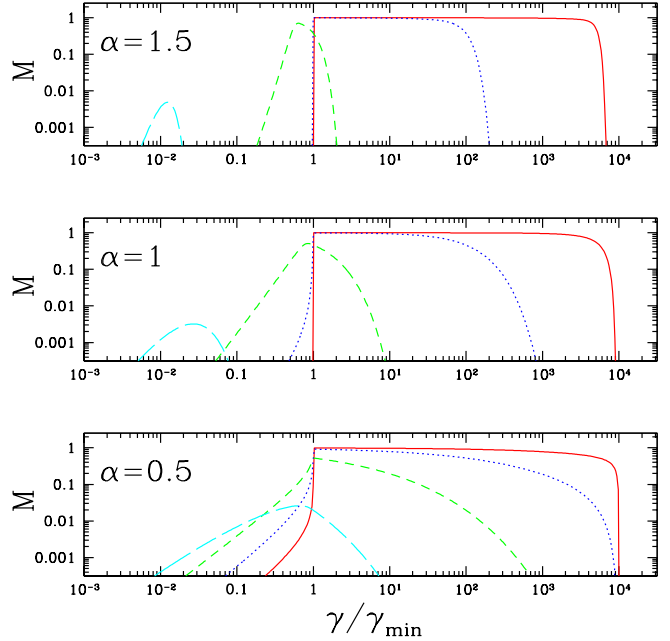}}
\caption{\protect\label{modul}
The modulation function $M(\ggam,\rho_{\rm d},p,\td)$ defined in 
Eq.~(\protect\ref{modulation}) as a function of $\ggam$, for 
$\rho_{\rm d}=10^{-4}$, $p=2.23$ and for $\alpha=1$ (diffusion), $\alpha=0.5$
(sub-diffusion) and $\alpha=1.5$ (supra-diffusion). In each case the solid, dotted, dashed and long-dashed curves
correspond to $\td=10^{-4}$, $10^{-2}$, $1$ and $10^2$, respectively.}  
\end{figure}

The effect of cooling in the relatively strong magnetic field of the head
region is described by the modulation function $M$ of 
Eq.~(\ref{modulation}). This function is plotted in Fig.~\ref{modul} for
various values of the ratio of diffusion to cooling time $\td$.

The character of the three different transport regimes is clearly seen from
this figure. In the case of supra-diffusion, $\alpha>1$, the effects of
cooling on the distribution of 
particles injected at the termination shock are very similar to
those calculated by \cite{kardashev62} under the assumption that each particle
spends the same length of time within the cooling zone. The upper cut-off
moves towards lower Lorentz factors as $\td$ rises, but remains quite sharp. 
A low energy tail at Lorentz factors less than those injected is almost absent
for $\td<1$, where the cooling time of the lowest
energy injected particle is longer than the 
transport time $\tdiff$. For larger values
of $\td$, the particles escaping from the head region have a sharply
peaked distribution centred at $\ggam\approx1/\td$. 

For diffusive transport ($\alpha=1$) the spread in escape times is appreciable,
resulting in a smoother cut-off and earlier development of a low energy tail
at Lorentz factors $\ggam<1$. Results for this case have been 
presented by \citet{wilson75} and \citet{valtaoja84}. In sub-diffusion 
($\alpha<1$) these trends are even stronger. Even for the relatively fast
transport through the cooling region implied by $\td=10^{-2}$, the particle
spectrum is steepened significantly over the entire injected range from 
$\ggam=1$ 
to $\ggam=10^4$. At low Lorentz factors a power-law tail is noticeable
already at $\td=10^{-4}$. The very large spread in escape times implied by this
model can be seen from the distribution plotted for $\td=100$. A
substantial number of particles leave the head region 
with Lorentz factors of 
$\ggam=10$. These particles have escaped within $10^{-3}$ of the 
characteristic transport time $\tdiff$. On the other hand, the distribution 
for $\td=10^{-4}$ 
reveals particles escaping with $\ggam=0.2$, which have
been retained in the head region for over $10^4\times\tdiff$.

In summary, the main effect of cooling in the high-loss head region 
on the particle
distribution is not to steepen it into a new power-law, but to impose
a cut-off of varying degrees of sharpness, according to the transport
properties. The adiabatic expansion losses suffered by these particles 
before entering the lobes shifts the escaping spectrum in energy
and amplitude, but does not
change its form.

\subsection{The lobe}

The effect of synchrotron and inverse Compton losses that are constant in time 
on a power-law distribution $\qlobe\propto\gamma^{-s}$
of injected electrons is well known.
However, losses in combination with time-dependent
adiabatic expansion and continuous injection --- 
processes present in Eq.~(\ref{diflobe}) --- considerably complicate 
this simple picture. Analytic solutions can be found 
for the regimes in which one or other of these processes dominates. Two of 
these
are given in the Appendix and are briefly discussed in this section. 
In each case it is assumed that a
power-law distribution is continuously injected in the range $\gone<\gamma<\gtwo$.

\subsubsection{Adiabatic losses}
In this case Eq.~(\ref{gdot}) becomes $\dot{\gamma} = -b\gamma/t$.
The solution of Eq.~(\ref{diflobe}) is 
given by Eq.~(\ref{asol1}). It is characterised by a range over which the
distribution has the same slope as the injection function $N\propto
\gamma^{-s}$ and accumulates approximately linearly with time. Above 
a critical value of the Lorentz factor, which decreases with time, the
distribution cuts off, tending to zero at $\gtwo$.
Below $\gone$, a power-law distribution is established with 
$N\propto -1+1/b$ which also accumulates approximately linearly with time.
(In the cases of relevance here, $b\approx 0.4$.)
The distribution vanishes below a lower cut-off, that
decreases with time.

\subsubsection{Inverse Compton losses}
In this case Eq.~(\ref{gdot}) becomes $\dot{\gamma} = -b_{\rm ic}\gamma^2$,
and the solution  of Eq.~(\ref{diflobe})
is given in Eq.~(\ref{csol1}). 
Below a time-dependent break frequency the distribution 
has the same slope as the injection function $N\propto\gamma^{-s}$ and
accumulates steadily with time. Above the break frequency the distribution 
steepens by unity: $N\propto \gamma^{-s-1}$ and remains constant in time,
finally cutting off at the maximum injected Lorentz factor $\gtwo$. For
$\gamma<\gone$ a cooling distribution $N\propto\gamma^{-2}$ is established.

\begin{figure}
\resizebox{\hsize}{!}{
\includegraphics{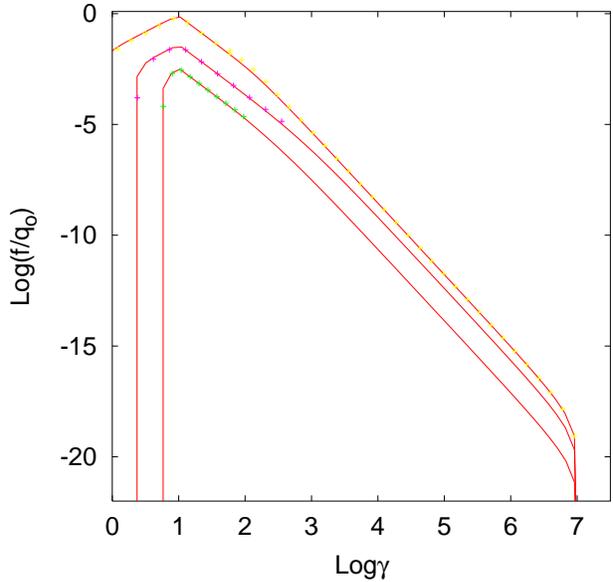}}
\caption{The evolution of the 
electron distribution in the lobe of a 
high jet power high 
redshift radio source, 
in the case of a power-law injection spectrum from the 
hotspot. From bottom to 
top: $t=1$, $10$ and $200\,$Myr, respectively. 
Continuous lines 
represent the numerical
results and crosses the analytical 
solutions as described by Eqs.~(\protect\ref{asol1}) 
and (\protect\ref{csol1}).}
\label{dslfig}
\end{figure}

Examples of these properties can be seen in Fig.~\ref{dslfig}, which 
shows the evolution of the normalised electron 
distribution, $N(\gamma,t)$ for the case of a high jet-power, high redshift 
radio source. The
continuous lines represent the numerical results and the crosses (where 
available) the analytical
solutions, as described in Eq.~(\ref{asol1}) and (\ref{csol1}). 
The values of the parameters
used in this case are: jet power $Q_0=1.3 \times 10^{40}\,$W, source redshift 
$z=2$, hotspot radius
$R=2.5\,$kpc, $\gone=10$ and $\gtwo=10^7$. 
The timescale associated with these parameters is
$t_0=2.5 \times 10^5\,$yr and the source activity is assumed to start
at $t_{\rm start}=t_0$; our numerical results are insensitive to this choice.
The efficiency was chosen to be $\eta=10^{-1}$. 

The amplitude of the 
energy spectrum increases with time since the rate at which particles 
enter the lobe is 
higher than the rate at which they 
cool to lower energies. 
At epochs earlier than 
$(b_{\rm s}/b_{\rm ic})^{1/a}=t_{\rm crit}\approx21\,$Myr
synchrotron losses dominate over Compton losses
at all energies. Adiabatic losses dominate below a critical Lorentz
factor given by 
$\gamma_{\rm crit}=bt^{a-1}/b_{\rm s}\approx10^2[t({\rm Myr})]^{0.57}$
at early times ($t<t_{\rm crit}$) and by
$\gamma_{\rm crit}=b/b_{\rm ic}t\approx10^4/t({\rm Myr})$ 
at late times ($t>t_{\rm crit}$). Thus, in 
Fig.~\ref{dslfig}, the distribution at $1\,$Myr is 
dominated by adiabatic losses for $\gamma<10^2$ and by (time dependent)
synchrotron losses for $\gamma>10^2$. At epochs $100\,$Myr and $200\,$Myr
Compton losses dominate above $\gamma=100$ and $50$ respectively.
Because the synchrotron losses decrease with time, even the loss-dominated
part of the distribution grows significantly between $1$ and $10\,$Myr. 
When Compton losses are decisive, 
the loss dominated part remains almost constant.
Below $\gamma=10$ (the minimum Lorentz factor at which injection occurs) 
a power-law distribution of index $-1+1/b\approx1.6$ is established by
adiabatic losses.

\section{Results}

In this section we present some results concerning the 
spectral evolution of the 
lobes of classical
double radio sources, derived by integrating the electron energy 
distribution in the lobes,
$N(\gm,t)$, over the electron emissivity given in Eq.~(\ref{power}). The 
electron energy
distribution is computed numerically by integrating Eq.~(\ref{odef}) 
using the
differential rate at which electrons are injected into the lobe 
given in Eqs.~(\ref{qlobea}) and (\ref{modulation}). 

The parameters we use to specify a radio source are:
\begin{enumerate}
\item
The jet power $Q_0$ 
\item
The efficiency $\eta$ ($\le1$) 
with which particles are accelerated at the terminal
shock front, which we set to unity in case~A (without re-acceleration),
and to $0.4$ in case~B (with re-acceleration).
The specific power is linearly
proportional to $\eta$.
\item
The area $A$ of the terminal hot spot --- we take 
$A=\pi\times (2.5)^2\,{\rm kpc}^2$.
\item
The redshift $z$ of the source
\item
The density of the surrounding matter at a fiducial distance, according to 
Eq.~(\ref{denprof}).  
\item
The density profile, for which we take $\beta=1.5$.
\item
The power-law index $p$ of particles accelerated at the terminal shock. Depending
on the Lorentz factor of the jet, one expects $p\approx2$ to $2.3$.
We adopt the value $2.23$ \citep{kirketal00}.
\item
The minimum Lorentz factor of the accelerated particles. This should be
somewhat larger than the Lorentz factor of the jet. The results are not
sensitive to its value for the relevant range of spectral indices.
\item
The ratio $\td$ of the typical transport time 
$\tdiff$ to cooling time $\tcool$ at $\gamma=\gmin$. 
This is essentially a free parameter. Although spherical geometry is used to 
motivate the transport model in Sect.~\ref{accel}, 
the resulting concept of a stochastic spread in particle escape
times from the high magnetic field head region is much more generic.
However, as in the case of standard diffusion, the model assumes infinite 
velocity of a fluid element within the head region.
This could lead to unphysical results if $\tdiff$ is chosen to be too small.
In our case, if $R_{\rm esc}$ is the minimum distance a fluid element must
travel before exiting the head region then the 
ratio of escape time to $\tcool$ 
must exceed $10^{-4}\gmin (Q/10^{40}\,{\rm W})(R_{\rm esc}/1~{\rm kpc})$
for all particles.
In the case of sub-diffusion, where the individual escape times have a large
spread about the mean, $\td$ must exceed this value by a factor of a few.
Thus, in our computations we take $\td\ge 2\times 10^{-3}$.   
\item
The maximum Lorentz factor $\gmax$ 
to which particles are accelerated at the terminal
shock. In accordance with X-ray observations of hot spots we assume
$\gmax\approx10^7$. As a result, $\td\gg\gmin/\gmax$ and 
the results are
insensitive to this parameter.
\item
The index $\alpha$ that characterises the transport process through the head
region: $0<\alpha<2$. We consider three values: sub-diffusion $\alpha=0.5$,
diffusion
$\alpha=1$ and supra diffusion $\alpha=1.5$.
\end{enumerate}

In addition we distinguish two cases:
\begin{itemize}
\item
Case A: particles suffer time-dependent adiabatic and radiative
losses on passing from the
termination shock to the lobes.
\item
Case B: adiabatic losses between the termination shock and the lobes are
compensated by re-acceleration in the head region. In this case $k(t)\equiv1$
in Eq.~(\ref{qlobea}).
\end{itemize}

\begin{figure}
\resizebox{\hsize}{!}
{\includegraphics{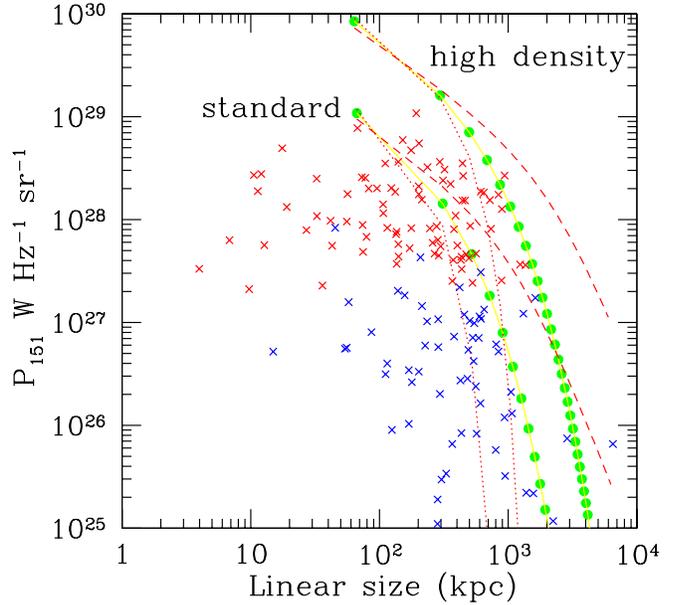}}
\caption{\protect\label{pdcasea}
Tracks for two high redshift, high power sources for case~A 
(no re-acceleration), with parameters as 
given in Table~\protect\ref{table1}. The \lq 
standard\rq\ track corresponds 
to \protect\citet{blundelletal99}, Fig.~13.
On each track the dots indicate the source age, starting at $1\,$Myr
and uniformly spaced in intervals of $5\,$Myr. 
The solid line shows the track for diffusion ($\alpha=1$) the dashed line for
sub-diffusion ($\alpha=0.5$) and the dotted line for supra-diffusion
($\alpha=1.5$). 
Crosses and open circles indicate FR~II sources of
the 3CR revised sample, with $z>0.5$ and $z<0.5$, respectively.}
\end{figure}

We consider first tracks on the power versus linear size diagram of
high power, high redshift sources. 
Figure~\ref{pdcasea} shows two such tracks for
case~A --- no re-acceleration of particles after the termination
shock --- for parameters given in Table~\ref{table1}.
Three different transport models are shown ($\alpha=0.5$, $1$, and $1.5$)
and we have chosen $\td=2\times10^{-3}$.
This figure also shows the positions of FR~II 
sources in the 3CR revised catalogue,
\citep{laingetal83} according to the classifications by 
\citet{jacksonrawlings97}, as updated and used by \citet{willottetal99} and 
published on-line by \citet{willott02}.
The linear size was computed assuming the cosmological parameters
$q_0=0$, $H_0=50\,{\rm km\,s^{-1}\,Mpc^{-1}}$. Sources with 
redshift $z>0.5$ are indicated by a cross, those with $z<0.5$ by 
an open circle.  

The first track, labelled \lq standard\rq\
can be directly compared with Fig.~13 of \citet{blundelletal99}, to reveal
the new features of our model.
We find a much sharper
decline in the source power with increasing radius than
predicted by \citet{blundelletal99}.
For these computations we adopt the minimum plausible value of 
the transport time: $\td=2\times10^{-3}$ --- increasing $\td$ always leads to 
an even steeper decline. 
The track passes well
below the large ($D>600\,$kpc) 
sources with specific power $P_{151}>3\times10^{28}
{\rm W\,Hz^{-1}}$ and decays to $P_{151}\approx7\times10^{26}
{\rm W\,Hz^{-1}}$ within less than $26\,$Myr.
The reason for this behaviour is the lack of particles of high Lorentz factor
entering the lobes. Synchrotron losses in the head region effect a cut-off of
the injected distribution in our model of acceleration and 
transport. After suffering the strong expansion 
losses associated with the pressure
difference between primary hot spot and lobe, this cut-off affects the number 
of particles available to radiate at $151\,$MHz, which, as a result, 
declines steeply with age.  
In order to account for the larger sources, it is necessary not only to assume
a much larger power $Q_0\approx6.5\times10^{40}\,$W, but 
also a much higher
surrounding density $\rho\approx10^{-22}{\rm kg\,m^{-3}}$
at $10\,$kpc. 
Such a track is
shown in Fig.~\ref{pdcasea} labelled \lq high density\rq. 
At age roughly $22\,$Myr this track coincides with the
largest high power source in the 3CR revised catalogue.
If sources existed with these extreme parameters they should also be
visible at much higher specific powers, when 
less than $10\,$Myr old. Such
objects, however, do not appear in the catalogue. We therefore conclude that
case~A tracks, in which particle acceleration occurs only at the termination
shock are in conflict with the data.

This conclusion is unaffected by the type 
of transport present in 
the head region. The dashed lines in Fig.~\ref{pdcasea} show the
effects of sub-diffusion and the dotted lines that of supra-diffusion
on the \lq standard\rq\ and \lq
high density\rq\ tracks. For ages greater than $5\,$Myr, sub-diffusion
always results in greater specific power than diffusion ($\alpha=1$),
whereas supra-diffusion gives a lower specific power. 
This arises because sub-diffusion produces a greater spread in escape times
from the head region than does diffusion. 
The upper cut-off in the Lorentz factor of
particles entering the lobes is correspondingly higher, whereas the
opposite is the case for supra-diffusion.
Nevertheless,
the sub-diffusion track for the standard source 
is unable to account for the largest high
redshift sources, and the supra-diffusion track for a high density
source would predict more medium-sized high specific power sources
than are observed.

\begin{figure}
\resizebox{\hsize}{!}
{\includegraphics{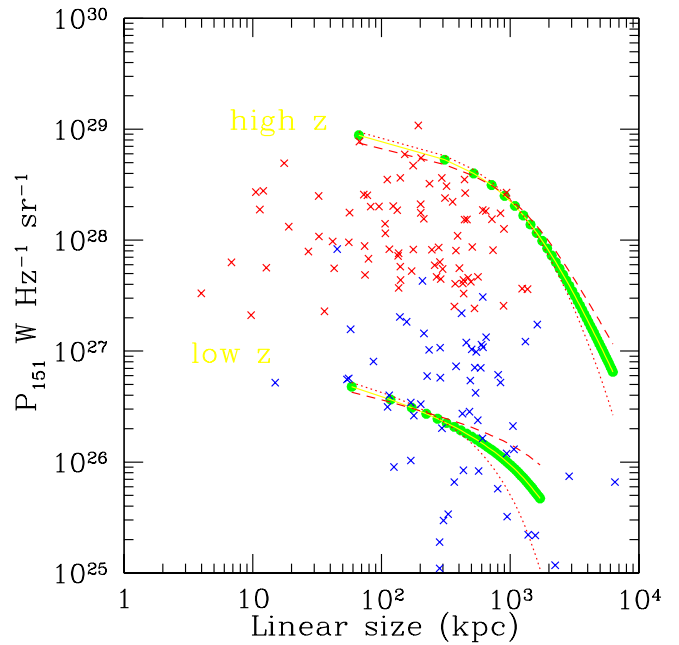}}
\caption{\protect\label{pdcaseb}
Tracks for a high redshift, high power source and 
a low redshift, low power source for case~B (including re-acceleration), 
with 
parameters as given in Table~\ref{table1}.
Diffusive transport $\alpha=1$ is shown by a solid line, on which the source
age is indicated by the dots, starting, for the high $z$ source, at
$1\,$Myr and, for the low-$z$ source, at $4\,$Myr and uniformly 
spaced in intervals of $5\,$Myr in each case. 
The dashed lines indicate sub-diffusion $\alpha=0.5$ and the dotted lines
supra-diffusion ($\alpha=1.5$)
Crosses and open circles indicate FR~II sources of
the 3CR revised sample, with $z>0.5$ and $z<0.5$, respectively.}
\end{figure}

Turning to case~B, in which the adiabatic losses between head region and lobe
are compensated by a re-acceleration process operating downstream of the
termination shock, we present in Fig.~\ref{pdcaseb} two tracks in the $P$--$D$
diagram, one 
corresponding to a high redshift, high power source and one to a low
redshift, low power source. Once again, the parameters are chosen to enable
direct comparison with the work of \citep{blundelletal99} and are shown in
Table~\ref{table1}. The value of $\td$ is the same as in Fig.~\ref{pdcasea}.

\begin{table}
\centering
\caption[ ]{\protect\label{table1}
Parameters of the models displayed in Figs.~\protect\ref{pdcasea}
  and \protect\ref{pdcaseb}} 
\begin{tabular}{l r@{.}l r@{.}l r@{.}l r@{.}l}
 &\multicolumn{2}{c}{$Q_0\,$(W)}&
\multicolumn{2}{c}{$\eta$} &
\multicolumn{2}{c}{$z$} &
\multicolumn{2}{c}{$\rho\,{\rm (kg\,m^{-3})}$}\\
 &\multicolumn{2}{c}{ }& 
\multicolumn{2}{c}{ }& 
\multicolumn{2}{c}{ }& 
\multicolumn{2}{c}{at $10\,$kpc }\\ 
\\
Fig.~\ref{pdcasea}, \lq standard\rq\
& 1&3$\times10^{40}$ &1&& 2&& 1&7$\times10^{-23}$\\
 \\
Fig.~\ref{pdcasea}, \lq high density\rq\
& 6&5$\times10^{40}$ &1&& 2&& 1&$\times10^{-22}$\\
 \\
Fig.~\ref{pdcaseb}, \lq high $z$\rq\
& 1&3$\times10^{40}$ &0&4& 2&& 1&7$\times10^{-23}$\\
 \\
Fig.~\ref{pdcaseb}, \lq low $z$\rq\
& 1&3$\times10^{38}$ &0&4& 0&2& 1&7$\times10^{-23}$\\
\end{tabular}
\end{table} 

A prominent feature of these tracks is the steepening as the source ages.
To a lesser extent, this is also seen in case~A, Fig.~\ref{pdcasea},
and is present in the tracks computed by \citet{kaiseretal97} and 
\cite{blundelletal99}. It arises at the point where 
inverse Compton losses become 
important in the lobes. Whereas the synchrotron
losses continually weaken with time, the inverse Compton losses remain
constant, causing a more dramatic fall off in the specific power.

\begin{figure}
\resizebox{\hsize}{!}
{\includegraphics[bb=18 146 582 700]{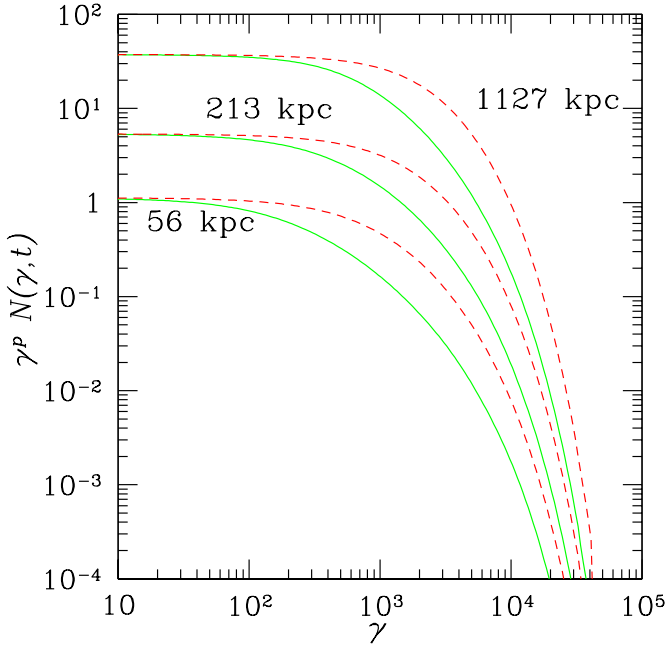}}
\caption{\protect\label{speccompare}
The particle distribution function in the lobes, multiplied by $\gamma^p$,
where $p$ is the index of the injected distribution for the high and low
redshift sources shown in Fig.~\protect\ref{pdcaseb}.  
The solid lines depict the distribution for the high $z$ source, the dashed
lines for the low $z$ source. For each source the distribution is plotted 
when it achieves the sizes 56, 213 and $1127\,$ kpc. For fixed size, the
higher redshift source has the softer spectrum. For fixed redshift, the
older (larger) source is softer.}
\end{figure}

Because of the dependence of the source size on its power 
(see Eq.~(\ref{ssize}))
the low $z$ source in Fig.~\ref{pdcaseb} is systematically older at a fixed
source size than is the high $z$ source --- a potentially important fact
in the 
interpretation of the correlation between source size and spectral index.
In Fig.~\ref{speccompare} we plot the particle 
distribution in the lobes, multiplied by $\gamma^p$, against Lorentz factor
$\gamma$ for both the high $z$ source (solid line) and the low $z$ source
(dashed line). Despite the age difference, it can be seen that where 
cooling steepens the spectrum, the high power, high $z$ source 
always has a spectrum that is softer than that of the low power, low $z$
source. Furthermore, for fixed redshift, the source spectrum systematically
softens with increasing size, and, therefore, also with increasing age.

\section{Conclusions}

We have introduced a new description of the acceleration and transport of
particles in the hot spots and lobes of FR~II radio galaxies. 
Using the standard approach of assuming a power-law spectrum injected at the
termination shock, we model the subsequent transport through the head region
into the lobes using a formalism which permits the investigation of 
\lq anomalous\rq\ regimes.
These transport 
regimes are described by the single parameter $\alpha$, which determines the
time-dependence of the mean-square displacement of a particle: 
$\left<\Delta r^2\right>\propto t^\alpha$. Standard 
diffusion, in which the flux is proportional to the gradient of the
particle density (Fick's law) corresponds to $\alpha=1$. The anomalous regimes
of sub-diffusion and supra-diffusion correspond respectively to $\alpha<1$ and
$\alpha>1$.
In the current application, the most important 
physical difference between these regimes is that they
permit a range of escape times for the particles from the high-loss head
region. Compared to the case of standard diffusion, these times 
have a much wider distribution
in sub-diffusion and a much tighter one in 
supra-diffusion. 
The new formalism also 
requires a parameter $\td$ that corresponds to the
average residence time in units of the cooling time.
Thus, in total, one more parameter is used 
than the standard 
\citet{kardashev62} model with a single escape time for all particles 
and one parameter fewer is needed than the model introduced by 
\citet{blundelletal99}, 
which requires two break frequencies and a \lq leakiness\rq.

The self-similar description of the hydrodynamical evolution of classical
double radio sources implies that the lobe pressure falls with time. 
Because the pressure in the primary hot-spots does not appear to undergo this
evolution, 
the particles injected at there in 
older sources have to overcome ever more severe adiabatic losses.
These losses were not modelled by \cite{kaiseretal97}, who specified the
distribution of particles entering the lobes. \citet{blundelletal99} pointed
out their importance and included them in their $P$--$D$ tracks. However,
because the transport model did not include an upper cut-off of injected
particles, but merely a spectral break, they were still able to account 
for the observed population of sources. We find that the adiabatic losses
between primary hot spot and lobe prevent this when our 
transport model is adopted. 
The large, high luminosity sources can be explained
only as very powerful jets in very dense environments, in which case 
they should have smaller, brighter predecessors, that are absent from the 3CR
revised catalogue.
There could be several reasons for this. One possibility is that 
at high redshift, small sources with powerful jets exist, but are 
not bright in the radio because of absorption. The missing progenitors could
then be powerful, high redshift analogues of the \lq\lq Gigahertz Peak 
Spectrum\rq\rq\ sources \citep{bicknelletal97}.
Another is that the jet power $Q_0$ remains quite weak during the early
evolution of a high-redshift source, becoming strong 
only after $\sim10\,$Myr. Intermittent jet activity in FR~II sources
has already been proposed
\citep{reynoldsbegelman97}, but as a solution to the opposite problem of an
observed {\em over}-abundance of small powerful sources (at lower redshift). In
general, it seems easier to imagine the jet power to be larger during 
earlier phases
of a source's life, leading to a faster decay of the 
specific source power with time.

However, the
solution we propose to this problem is that acceleration is not confined
to the primary hot spot, but occurs throughout the head region. 
Independent evidence from individual sources in the infra-red, 
optical and X-ray bands
\citep{meisenheimeretal96,meisenheimeretal97,perleyetal97,wagnerkrawczynski00}
supports this idea.
A plausible
mechanism is repeated encounters with weaker shock fronts which may permeate
this turbulent region. We have investigated the $P$--$D$ tracks and spectra of
a model in which it is assumed that this distributed acceleration compensates
the expansion losses between primary hot spot and lobes, without significantly
modifying the spectrum. Our tracks are more akin 
to those presented by \cite{kaiseretal97} and we have investigated their 
sensitivity to the new transport parameter $\alpha$. 
We have shown that some of the properties noted
by \citet{blundelletal99}, such as a redshift/spectral index and a 
size/spectra index correlation also emerge from our model.

\begin{acknowledgements}
We thank Klaus Meisenheimer and Stefan Wagner for helpful discussions and 
Jane~Dennett-Thorpe for a careful reading of the manuscript and 
valuable suggestions. 
K.M. acknowledges support from the German Research Council (DFG)
under SFB~439. 
\end{acknowledgements}

\newpage
\onecolumn
\section{Appendix}
Analytic solutions of Eq.~(\ref{diflobe}) are presented for the cases 
in which either only adiabatic losses or only synchrotron losses are present,
assuming an injection function given by
\beq
\qlobe(\gamma)=
\left\lbrace \begin{array}{ll}
q_0 \gamma^{-s}& \textrm{for $\gone \le \gamma \le \gtwo$ } \\
        &       \\
0 & \textrm{otherwise}
\end{array}
\right.             
\label{sourceappendix}
\eeq

\subsection{Adiabatic losses}
The solution of Eq.~\ref{diflobe}, provided that $\gone < 
\gtwo(t/{\tstart})^{-
b}$, is:
\begin{equation}
f(\gamma,t)= \left\{ \begin{array}{ll}  

0,& \mbox{$\gamma<\gamma_{a_1}$}\\
        &       \\
{q_0 {\tstart} \gamma^{-s} \over
\lambda}\left({\gamma \over \gone}\right)^{\lambda\over b} 
\left({t \over {\tstart}}\right)
\left\lbrace 1- {\left[\left({\gamma \over \gone}\right) 
\left({t \over {\tstart}}
\right)^b\right]}^{-{\lambda \over b}}\right\rbrace,
        & \mbox{$\gamma_{a_1} \le \gamma \le \gone$}\\
        &       \\
{q_0 {\tstart} {\gamma}^{-s} \over
\lambda }\left({t \over {\tstart}}\right)
\left[1-\left({t \over {\tstart}}\right)^{-\lambda}\right], & \mbox{$\gone< 
\gamma<\gamma_{a_2}$}\\
        &       \\                       
{q_0 {\tstart} {\gamma}^{-s} \over \lambda }\left({t \over
{\tstart}}\right)\left[1-\left({\gamma \over \gtwo}\right)^{\lambda 
\over b}\right], &
\mbox{$\gamma_{a_2} \le \gamma \le \gtwo$} \\
        &       \\
0, & \mbox{$\gamma > \gtwo$}
\end{array}
\right.
\label{asol1}
\end{equation}
with $\lambda=b(s-1)+1$ and $\gm_{a_1}=\gone({t /{\tstart}})^{-b}$, 
$\gm_{a_2}=\gtwo({t / {\tstart}})^{-b}$.

The solution of Eq.~\ref{diflobe}, for $\gone < 
[\gtwo^{-1} +b_{\rm ic}(t-\tstart)]^{-1}$, is:
\begin{equation}
 f(\gamma,t)= \left\{ \begin{array}{ll}
0,& \mbox{$\gamma< \gamma_{c_1}$}\\
        &       \\
K{\gamma}^{-(s+1)}
\left\lbrace \left({\gamma \over \gone}\right)^{s-1}
\left[1-b_{\rm ic} \gamma(t-{\tstart})\right]^{s-1}\right\rbrace, &
\mbox{$\gamma_{c_1} \le \gamma \le \gone$}\\
        &       \\
K{\gamma}^{-(s+1)}
\{1-[1-b_{ic}\gamma(t-{\tstart})]^{s-1} \}, &
\mbox{$\gone< \gamma < \gamma_{c_2} $}\\
        &       \\
K{\gamma}^{-(s+1)}\left[1-({\gamma \over
\gtwo})^{s-1}\right], &
\mbox{$\gamma_{c_2} \le \gamma \le \gtwo$}\\
        &       \\
0, & \mbox{$ \gamma > \gtwo$}
\end{array}
\right.
\label{csol1}
\end{equation}
where $K={q_0/ {b_{\rm ic}(s-1)}}, \gamma_{\rm c_1}
={[\gamma_{\rm 1}^{-1}+b_{\rm ic}(t-\tstart)]}^{-1}$ and $\gamma_{\rm
c_2}={[\gamma_{2}^{-1}+b_{\rm ic}(t-\tstart)]}^{-1}$.

In the case that adiabatic losses determine the spectrum of a 
source the 
solution of Eq.~(\ref{diflobe}) for $\gone > \gm_{a_2}$, is:

\begin{equation}
f(\gamma,t)= \left\{ \begin{array}{ll}

0,& \mbox{$\gamma<\gamma_{a_1}$}\\
        &       \\
{q_0 {\tstart} \gamma^{-p} \over
\lambda}\left({\gamma \over \gone}\right)^{\lambda \over b}
\left({t \over {\tstart}}\right)
\left\lbrace 1- \left[\left({\gamma \over \gone}\right) \left({t \over
{\tstart}}\right)^b\right]^{-{\lambda \over b}}\right\rbrace,
        & \mbox{$\gamma_{a_1} \le \gamma \le \gamma_{a_2}$}\\
        &       \\
{q_0 {\tstart} {\gamma}^{(1-b)/b} \over
\lambda }\left({t \over {\tstart}}\right) {\gone}^{-\lambda/b}
\left[1-\left({\gone \over \gtwo}\right)^{\lambda/b}\right], & \mbox{$\gamma_{a_2} <
\gamma \le \gone$}\\
        &       \\
{q_0 {\tstart} {\gamma}^{-p} \over \lambda }\left({t \over
{\tstart}}\right)\left[1-\left({\gamma \over \gtwo}\right)^{\lambda \over b}\right], &
\mbox{$\gone \le \gamma \le \gtwo$} \\
        &       \\
0, & \mbox{$\gamma > \gtwo$}
\end{array}
\right.
\label{asol2}
\end{equation}                 

In the case that inverse compton losses determine the spectrum of a 
source the 
solution of Eq.~(\ref{diflobe}) for $\gm > \gm_{c_2}$, is:

\begin{equation}
 f(\gamma,t)= \left\{ \begin{array}{ll}
0,& \mbox{$\gamma< \gamma_{c_1}$}\\
        &       \\
K{\gamma}^{-2}
\{ \gone^{1-p}-{[\gm^{-1}-b_{\rm ic}(t-{\tstart})]}^{p-1}\}, &
\mbox{$\gamma_{c_1} \le \gamma \le \gamma_{c_2}$}\\
        &       \\
K{\gamma}^{-2}
(\gone^{1-p}-\gtwo^{1-p}), &
\mbox{$\gamma_{c_2}< \gamma < \gone $}\\
        &       \\
K{\gamma}^{-(p+1)}\left[1-\left({\gamma \over
\gtwo}\right)^{p-1}\right], &
\mbox{$\gone \le \gamma \le \gtwo$}\\
        &       \\
0, & \mbox{$ \gamma > \gtwo$}
\end{array}
\right.
\label{csol2}
\end{equation}          

\twocolumn

\end{document}